# MyTm: An Automated Melting Temperature Calculation Toolkit

Yu S. Huang,[1] Hong X. Song,[1] Y. Sun,[1] Jin L. Li[1], Yin L. Xu[1], F. C. Wu[1], Y. C. Gan[1], Yu. F. Wang,[1] Hao Wang,[1*] Hua Y. Geng[1,2†]

*1 National Key Laboratory of Shock Wave and Detonation Physics, Institute of Fluid Physics,*

*China Academy of Engineering Physics, Mianyang, Sichuan 621900, P. R. China;*

*2 HEDPS, Center for Applied Physics and Technology, and College of Engineering, Peking University, Beijing 100871, P. R. China.*

* To whom correspondence should be addressed. E-mail: wh_95@qq.com

† To whom correspondence should be addressed. E-mail: s102genghy@caep.cn

**ABSTRACT:** Melting temperature calculation is one of the important topics in computational materials science. In high-throughput in silico screening and artificial intelligence assisted design of materials, it usually requires a rapid and autonomous assessment of the melting temperature of the target. Unfortunately, molecular dynamics (MD) simulations of the melting point require many cumbersome and manual operations, making large-scale calculation of the melting point challenging. In this work, we introduce *MyTm*, a toolkit that employs MD to automatically determine the melting point. The method is fully modularized, and by combining these modules, the program enables fully automated melting calculations by using commonly adopted approaches, including the direct-heating method, the void method, the modified void method, the solid–liquid coexistence method, and the Z method. Moreover, a machine learning (ML) method is proposed employed to recognize and classify the solid-like and liquid-like atoms, which effectively resolve the low accuracy issue in conventional classification approaches, thus making the automated high-throughput pipeline of melting-point calculation possible. The robustness and efficacy of *MyTm* have been demonstrated by several well-studied systems.

**PROGRAM SUMMARY**

*Program title:* MyTm

*CPC Library link to program files:* (to be added by Technical Editor)

*Developer's respository link:*

*Licensing provisions:* MIT

*Programming language:* C

*External routines/libraries:* GSL [1], Intel MKL [2]

*Nature of problem:* A fully automated implementation of direct molecular dynamics simulations for melting point determination is developed to eliminate the labor-intensive procedures involved in conventional melting point calculations and to facilitate the high throughput materials screening. Machine learning is employed to enhance the identification of solid and liquid atomic environments during the melting point calculation process.

*Solution method:* First, the direct molecular dynamics calculation of the melting point is decomposed into four general sub-processes: solid and liquid atomic environments classification, construction of the melting model, ensemble selection, and temperature control strategy. Second, full coverage of molecular dynamics–based melting point calculation methods, including the direct-heating method, the void method, the modified void method, the two-phase method, the sandwich method, the Z method, and the modified Z method [3–8], is achieved through the coordinated integration of these four submodules. Third, machine learning and related approaches are employed to improve the accuracy of automated solid–liquid atomic environment identification, thereby enabling fully automated melting point determination.

*Additional comments including Restrictions and Unusual features:* The program enables automated execution of all direct molecular dynamics–based melting point calculation methods, but it does not support melting point determination via free-energy approaches. In addition, a small-cell method is integrated to accommodate scenarios with limited computational resources.

# I. Introduction

Melting temperature, or melting point is one of the most fundamental properties of materials, and plays a crucial role in the study of phase diagrams [1]. High-throughput melting point estimation is of great significance for accelerating the development of new materials, understanding the nature of matter, and validating or optimizing theoretical models[2–4]. At present, first-principles approaches for calculating the melting point of materials mainly fall into two categories: (i) the free-energy method; and (ii) the direct simulation method by using molecular dynamics (MD).

The free energy method is based on the thermodynamic criterion that the Gibbs free energies of the solid and liquid phases are equal at the melting point. It is typically implemented via thermodynamic integration, Clapeyron equation, or construction of thermodynamic surfaces. While generally considered highly reliable, its results are highly sensitive to the choice of reference model, integration path, and sampling quality, and its high computational cost limits its application in high-throughput calculations.[5–7] Currently, the state-of-the-art approach for quick melting-point evaluation is direct MD simulations. The main advantage is that, compared to the free-energy method, direct simulation method requires substantially fewer computational resources. However, due to the limitations such as the size effect and superheating issue, the accuracy is generally lower than that of free-energy approaches.[8] Overall, this method is more preferred for rapid assessments of melting points in scenarios where a high level of accuracy is not required. Such commonly used methods include: direct-heating method, void method, modified void method, the solid–liquid coexistence method, sandwich method, Z method, and modified Z method.[9–14]

Since the melting criterion in most direct molecular dynamics methods is ultimately determined from the evolution of solid and liquid phases, reliable identification of solid–liquid structures is essential for automated workflows. Numerous order parameters have been proposed for atomic structure identification, including, order parameter reported by Steinhardt et al.[15] for the study of glassy structures, order parameter designed by Chau and Hardwick for tetrahedral configurations[16] and the polyhedral template matching method.[17] In recent years, machine learning–based atomic structure identification has developed rapidly.[18,19] For example, Ziletti et al. employed a multilayer perceptron (MLP) network to automatically classify crystal structures according to their symmetry.[20] Scheiber et al. used an MLP network to classify binary salts, thereby enabling automated melting point calculations.[21] Zhao et al. developed an MLP network for predicting the crystal space groups of materials.[22] Building upon these structure identification techniques, several automated workflows for melting point calculations have been developed. For example, SLUSCHI[23] exploits the dynamics of the solid–liquid interface in finite-size systems, avoiding explicit free-energy calculations while reducing computational cost. Meanwhile, tools such as Pyiron and LAVA[24,25] provide the workflow of the solid–liquid coexistence method, significantly improving reproducibility and high-throughput capability. On the other hand, Dai et al. reformulate the melting problem as a temperature–volume fitting task, enabling a unified treatment of systems with and without elemental segregation.[26] Consequently, its applicability may be limited for systems exhibiting only a small solid–liquid volume difference. At present, these melting point calculation programs are essentially implementations of specific melting point calculation method. Therefore, developing an automated framework that transcends a single methodological paradigm, enables flexible combinations of multiple computational strategies and model constructions, and simultaneously improves the accuracy and robustness of solid–liquid phase identification remains an urgent challenge.

To address this issue, in this work, we modularize the melting-point calculation process by decomposing it into four key components: model construction, thermodynamic ensemble selection, temperature control, and solid–liquid classification (or solid fraction determination). A machine learning (ML) method that can automatically capture the solid–liquid structural features in terms of order parameters and atomic local densities is introduced, thus achieving accurate and efficient on-the-fly classification of solid and liquid atoms. These advances and modularized design enable full automation of melting-points calculation. All these functionalities are implemented in a C language based program named *MyTm*, the first for such tasks to our best knowledge.

## II. Features and Methodology

For direct MD-based melting point calculations (e.g., the solid–liquid coexistence method and the Z-method), the computational workflows reported in different studies exhibit consistent structural characteristics, a feature that is also reflected in existing automated implementations toolkit.[23–25] In general, such methods can be abstracted into several key steps: (1) construction of an initial model appropriate for the target method; (2) selection of a suitable MD ensemble to ensure physically meaningful system evolution; (3) identification of solid and liquid atoms during the simulation; (4) dynamic adjustment of the simulation temperature based on the step (3) to approach the melting point, where the latter two steps typically form an iterative convergence process. These common features enable different direct MD methods to be implemented within a unified modular programming framework illustrated in Fig. 1.

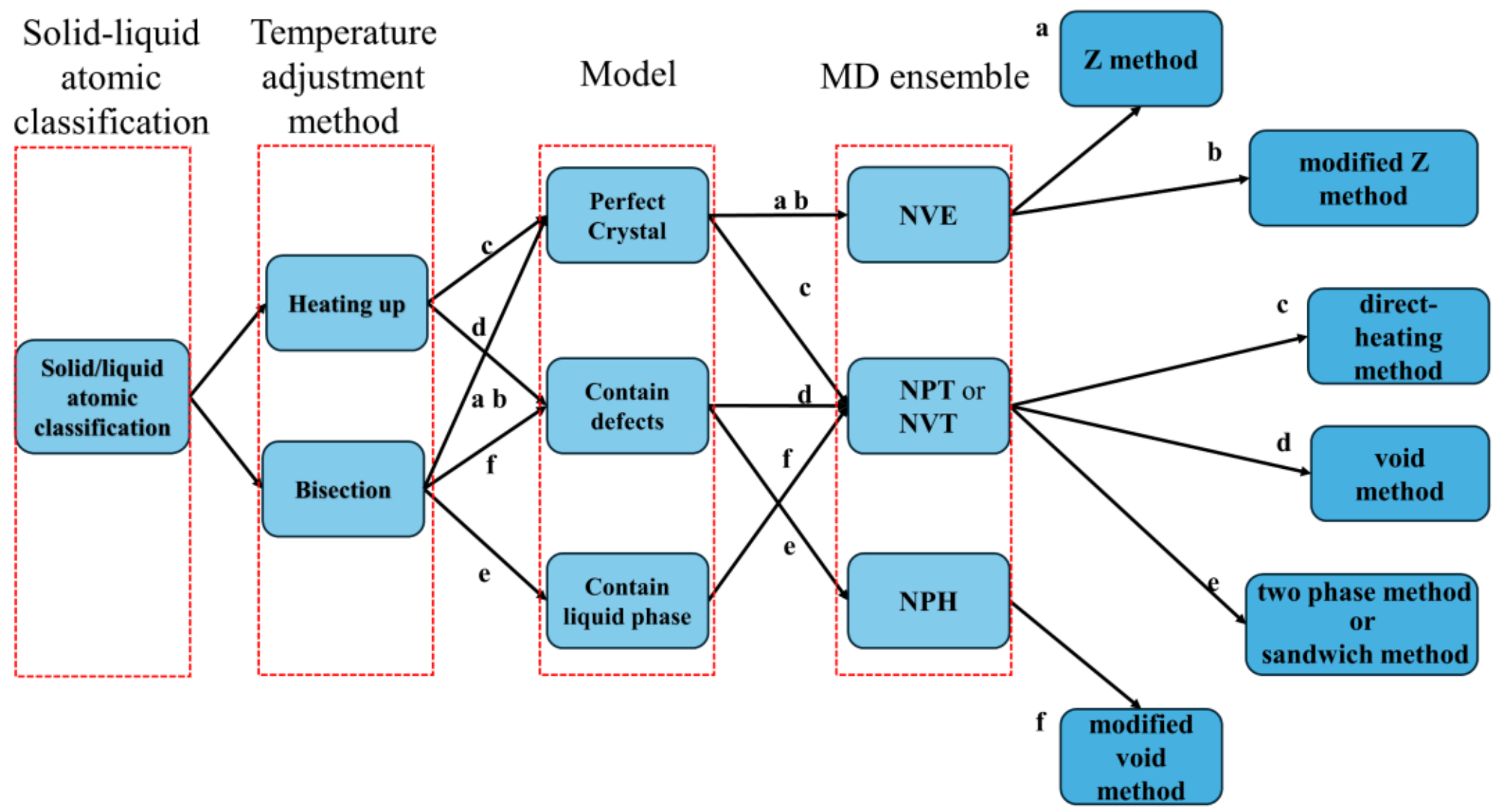

**Figure 1.** Constituent components of the melting point calculation methods as implemented in *MyTm* — different choice of the temperature adjustment method, model, and MD ensemble collectively determine the type of the melting point calculation method (the path is labeled by letter a-f, respectively).

### 1. Model construction

The structural models usually used for melting-point calculations mainly include perfect crystal model, defective crystal model, solid–liquid coexistence model, and hybrid model[8]. As shown in Fig. 2(a), the perfect crystal model is commonly employed in direct-heating and Z method.[27–29] This simple model is constructed by enlarging the simulation cell until finite-size effects are reduced to an acceptable level and reliable thermodynamic statistics can be obtained. However, such single-phase approach suffers from superheating, which leads to an overestimation of the melting temperature.[30] One way to address this issue is to use a defective crystal model (shown in Fig. 2 (b)), which is typically employed in the void method.[31] By introducing defects, the nucleation barrier of the liquid phase can be lowered, thereby reducing the error induced by superheating.[32] However, the melting point calculated from defective crystal models depends on the defect size and concentration, and therefore multiple tests are required to ensure reliable and accurate results.[33] Another type of melting-point model is the solid–liquid coexistence model.[11] Based on the theory of solid–liquid coexistence, this two-phase model (shown in Fig. 2(c)) has a clear physical foundation and provides more accurate melting-point results.[34,35] The construction of a solid–liquid coexistence model is relatively complicate: it comprises to build an elongated perfect crystal, fix half of the atoms, and then perform heating and annealing to obtain a solid–liquid coexistence structure.[36] Because the area of the solid–liquid interface affects the computational efficiency, multilayer coexistence models (i.e., sandwich method[37]) and spherical liquid-phase models are also frequently employed. In addition to these three types, hybrid models combining defects and liquid phase are sometimes used, which generally provide good accuracy but are the most complex to construct.

(a) Direct heating method, Z method and Modified Z method

supercell 10 10 10

perfect supercell

supercell 5 5 20

perfect elongated supercell

(b) Void method

supercell 10 10 10
build defect 2 2 0.5
size 0.8 0.8 0.8 0.5 0.5 0.5
pos 0.5 0.5 0.5 0.5 0.5 0.5
rot 0 0 0 0 0 0

layer defect

supercell 10 10 10
build defect 2 0 1.0
size 0.5 0.5 0.5
pos 0.5 0.5 0.5
rot 0 0 0

spherical defect

(c) Solid-liquid coexistence method

supercell 5 5 10
build liquid 1 0
size 1.0 1.0 0.5
pos 0.5 0.5 0.75
rot 0 0 0

solid-liquid coexistence model

supercell 5 5 10
build liquid 1 0
size 1.0 1.0 0.25
pos 0.5 0.5 0.625
rot 0 0 0
build liquid 1 0
size 1.0 1.0 0.25
pos 0.5 0.5 0.125
rot 0 0 0

sandwich method model

(d) Modified void method

supercell 5 5 20
build defect 1 0 0.5
size 1.0 1.0 0.1
pos 0.5 0.5 0.5
rot 0 0 0

void-containing elongated supercell

**Figure 2.** Schematic illustrations of the models used in different melting-point calculation methods, along with the corresponding construction commands implemented in *MyTm*. (a) Model employed in the direct-heating method, the Z method and the Modified Z method. (b) Model commonly used in the void method. (c) Models commonly adopted in the solid–liquid coexistence method. (d) Model for the modified void method.

## 2. Thermodynamic ensemble selection

The thermodynamic ensembles used in melting-point calculation are closely related to the underlying principles of each method. For direct-heating method and the void method[27], the NPT ensemble is typically employed, in which the temperature is gradually increased until the lattice melts. For solid-liquid phase equilibrium methods, the NPT or NPH ensembles are commonly recommended.[38,39] The NVT ensemble sometimes is also employed for these methods though. On the other hand, one should note that in the NPT/NVT ensemble with the supercell size that is accessible by first-principles method, long MD simulations at a fixed temperature usually leads the system to evolve into either a fully solid or a fully liquid state. By performing multiple simulations at different temperatures, the melting point can be approached iteratively. In contrast, in the NPH/NVE ensemble, the enthalpy (or energy) is controlled as constant.[40] The phase transformation between solid and liquid causes volume change and heat absorption or release, which naturally balance excessive heating or cooling. When the initial enthalpy is appropriately chosen, long simulations yield a stable solid-liquid equilibrium structure, with the system temperature maintained at the melting temperature. The NPH ensemble is usually employed for the modified void method , whereas the NVE ensemble is typically applied the Z method, where the system is allowed to naturally superheat and melt under constant total energy, enabling the determination of the melting point from the characteristic Z-shaped pressure–temperature relation.[41–43] Since *MyTm* operates by invoking external MD codes, the selection of thermodynamic ensemble must be specified by the user within the corresponding MD code.

## 3. Temperature adjustment

For direct MD–based melting points determination, there are essentially two temperature adjustment categories according to the purpose of the simulation. The first one involves continuously heating a crystalline system until it melts, which includes the direct-heating method and the void method. The second category aims to find an appropriate initial temperature (or equivalently, energy or enthalpy) that maintains a solid–liquid coexistence phase, which includes the solid–liquid coexistence method, the Z method, the modified Z method, and the modified void method. This strategy typically requires multiple MD simulations. The most common implementation is to divide the temperature into a series of grid points and progressively refine the grid to approach the equilibrium temperature. Another implementation employs a bisection scheme: one first identifies a temperature at which the system is clearly solid and another one at which it is clearly liquid, and then iteratively evaluates the midpoint temperature between the solid and liquid states to progressively bracket the equilibrium temperature. Because each iteration of the bisection method halves the temperature interval, it is highly efficient. *MyTm* adopts the bisection method in the second strategy. Therefore, from a program design perspective, the temperature adjustment in all melting point determination methods can be implemented by recombination of the sequential procedures that belong to two basic computational frameworks as shown in Fig. 3.

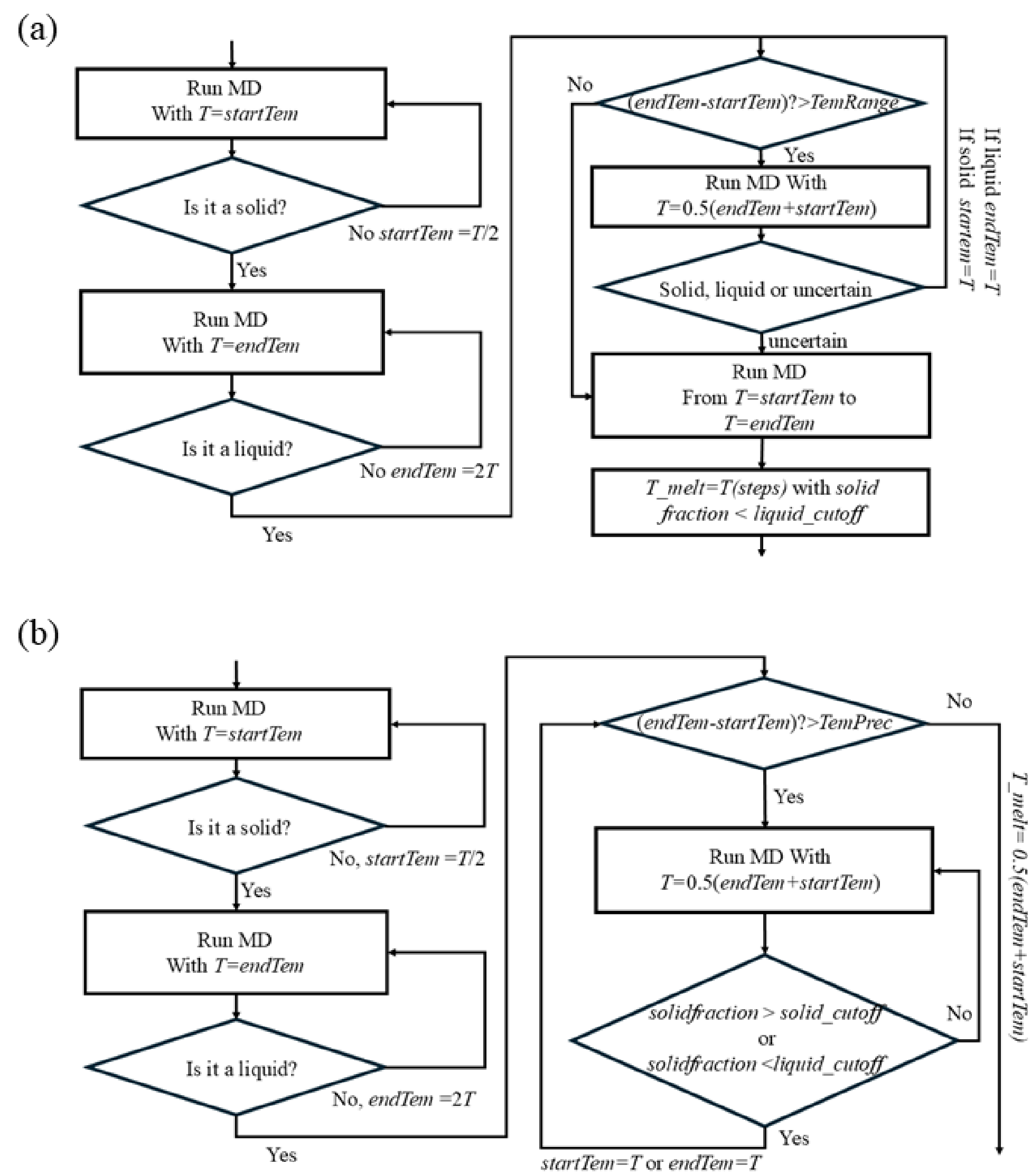

**Figure 3.** Flowchart of the temperature adjustment procedures in *MyTm*. (a) Direct heating method: The program first verifies and adjusts whether the low and high temperatures correspond to the solid and liquid phases, respectively. It then performs MD simulations according to the heating rate and detects the melting point. (b) Bisection method: The program first checks and adjusts whether the low and high temperatures correspond to the solid and liquid phases, respectively, and then iteratively applies the bisection approach between the two phases to approximate the melting point.

## 4. Solid-liquid atomic classification

Determining the melting fraction is one of the most challenge tasks in autonomous melting point calculations. In general, melting can be identified through discontinuities in certain thermodynamic quantities, such as temperature or pressure. However, in models that already contain a liquid phase—such as the modified void method or the solid–liquid coexistence method—this criterion becomes ineffective. Similar problem also appears in the traditional criteria that based on radial distribution function (RDF), mean square displacement (MSD) or atomic density. [44,45] In *MyTm*, Steinhardt order parameter (SOP) classification and atomic local atomic environment overlap (LAEO) classification are used. Furthermore, a ML technique are employed to improve the identification quality of solid and liquid atoms, as well as the useful technique of the maximum area under the curve.

### 4.1 Steinhardt order parameter classification

The SOP[15] is commonly used to identify the structure of simple elements, including body-centered-cubic (*bcc*), face-centered-cubic (*fcc*), hexagonal-close-packed (*hcp*). It was designed to be rotationally and translationally invariant by using spherical harmonics. Its expression is given by

$$q_l(i) = \sqrt{\frac{4\pi}{2l+1}\sum_{m=-l}^{l} \left|\frac{1}{N(i)}\sum_{j=1}^{N(i)} Y_{lm}(\boldsymbol{r}_{ij})\right|^2} \tag{1}$$

Where $N(i)$ represents the coordination number of the $i$th ion, and $Y_{lm}(r_{ij})$ denotes the spherical harmonics given by the ion pair (*i, j*). For the identification of solid and liquid phases, a variant of the standard Steinhardt SOP can be employed, which considers not only the central atom but also the coordinated atoms of those neighboring atoms, making the solid-liquid identification more accurate.[46,47] The expression of this SOP is given as follows:

$$s_l(i,j) = \sum_{m=-6}^{6} \left(\frac{1}{N(i)}\sum_{k=1}^{N(i)} Y_{lm}(\boldsymbol{r}_{ik})\right) \times \left(\frac{1}{N(j)}\sum_{k=1}^{N(j)} Y_{lm}(\boldsymbol{r}_{jk})\right)^* \tag{2}$$

$$\tilde{s}_l(i) = \frac{1}{N(i)} \sum_{j \epsilon N(i)} s_l(i,j)$$

### 4.2 Local atomic environment overlap order parameter classification

The LAEO classification is based on the concept of local atomic density in the SOAP[48] descriptor to distinguish between solid-like and liquid-like atoms.[49] It identifies the similarity between a target atom and a reference solid atom environment by performing an overlap integral of their respective local density functions. The expression is given as follows:

$$\rho_\chi(r) = \sum_{i\in\chi} \exp\left(-\frac{|r_i - r|^2}{2\sigma^2}\right) \tag{3}$$

$$\kappa_{\chi 0}(\chi) = \int \rho_\chi(r)\rho_{\chi 0}(r)dr \tag{4}$$

$$\tilde{\kappa}_{\chi 0}(\chi) = \frac{\kappa_{\chi 0}(\chi)}{\kappa_{\chi 0}(\chi 0)} = \frac{1}{n}\sum_{i\epsilon\chi}\sum_{j\epsilon\chi 0} \exp\left(-\frac{\left|r_i - r_j\right|^2}{4\sigma^2}\right) \tag{5}$$

Here, $\chi$ and $\chi 0$ represent the target atomic environment and reference atomic environment, respectively. $n$ is the normalization coefficient of function $\kappa_{\chi 0}(\chi 0)$.

### 4.3 Machine learning classification

For all classification methods mentioned above, including the SOP classification and the LAEO classification often yields unsatisfied performance in systems with complex coordination environments. For example, in the $MgSiO_3$ crystal the presence of neighboring oxygen atoms disrupts the rotational symmetry of the coordinated atoms, leading to poor accuracy of the SOP classification and causing some solid atoms to be misclassified as liquid ones. To address this challenge, *MyTm* also implements a multilayer perceptron (MLP)[50–54] model with three hidden layers $(15 \times 15 \times 15)$ for solid–liquid classification. The ML input vectors are represented by third-order neural equivariant potential[55] descriptor for local atomic environments. The MLP employs the leaky ReLU activation function[56], and the gradients are optimized using the Adaptive Moment Estimation (Adam) algorithm.[57]

The SOP and LAEO classifier outputs in [0, 1] are regarded as the probability of solid-like atomic environments. The MLP learns the mapping between atomic environments and their corresponding probabilities of being in the solid state. The learned probability values are subsequently combined through a weighted fusion, expressed as:

$$P_{out} = \frac{\alpha P_{SOP}}{\alpha P_{SOP} + \beta(1 - P_{LAEO})} \tag{6}$$

Here, $P_{out}$ denote the solid-state probabilities predicted by the MLP. The coefficients $\alpha$ and $\beta$ are the weights associated with the SOP and LAEO classifications. As a result, $P_{out}$ combines the SOP and the LAEO classification to yield more accurate results. This proposed ML classification has shown remarkable performance. Compared to the SOP classification and LAEO classification, the ML classification perfectly resolves the misclassification issues encountered in other methods. Furthermore, the solid–liquid interface becomes much clearer, and an accurate solid/liquid phase fraction.

## III. Implementation

### 1. Parameter control file

*MyTm* reads the *input.dat* file in the current directory as its control file. This file can be automatically generated by the *MyTm* as a template for the user to use or modify. The command to generate the *input.dat* file is

***$ MyTm -m method***

Here, ***method*** is an integer number representing the method for calculating melting point, and relevant information can be obtained through the ***-h*** command. The *Input.dat* file mainly contains four groups of commands corresponding to the four modules of *MyTm*: model construction, selection of the ensemble, temperature adjustment scheme, and solid–liquid atom identification.

### 1.1 Model construction parameters

*MyTm* has the build-in function to automatically construct the model, this includes expanding the size of the initial model, introducing defects, and generating solid–liquid coexistence phases. These model configurations are specified through corresponding tags in the *input.dat* file.

(i) Model expansion: The size of the simulation cell can be enlarged by setting the

parameters ***supercell*** in the *input.dat* file to generate a supercell. These parameters require three integers, which represent the multiplication factors of the simulation box along the three lattice vector directions, respectively. For example, to construct a supercell with a size of $2 \times 3 \times 4$, write

***supercell 2 3 4***

in *input.dat* file.

(ii) Defect and liquid-phase construction: The creation of defects or liquid regions is controlled by the ***build*** command in the *input.dat* file.

To construct a defect:

***build defect shape1 shape2 scale***

To construct a liquid phase:

***build liquid shape1 shape2***

Here, ***shape1*** and ***shape2*** are integer numbers (0: None, 1: Parallelepiped shape, 2: Ellipsoid shape, 3: Cylinder shape) representing different geometric configurations of the model, while ***scale*** is a float number between 0 and 1 that specifies the fraction of atoms to be removed when generating a defect. It should be noted that ***shape1*** defines the region inside which a liquid phase or defect is generated, whereas ***shape2*** defines the region outside which a liquid phase or defect is generated. The combination of these two tags allows for the construction of layered models. The parameter ***shape2*** is optional and may be set to 0. For example, a layered spherical defect is constructed in Fig. 2(b), in which setting ***shape1*** and ***shape2*** to 2 specifies ellipsoidal geometries, where ***shape1*** corresponds to a larger sphere and ***shape2*** to a smaller one. In this case, *MyTm* constructs defects inside the region defined by ***shape1*** and outside the region defined by ***shape2***, resulting in a shell-shaped spherical defect.

The ***build*** command can also include additional parameters specifying the central position (***pos***), bounding box size (***size***), and rotation (***rot***) of the ***shape1*** and ***shape2***. For the ***pos***, ***size*** and ***rot*** options, each is followed by three or six parameters, where the first three and the last three correspond to the properties of ***shape1*** and ***shape2***, respectively. The parameters specified by ***size*** and ***pos*** are expressed in fractional coordinates relative to the simulation box, while those of ***rot*** represent the rotation angles (in radians) around the x, y, and z axes, respectively. Multiple ***build*** commands can be employed in the *input.dat* file to construct a complex model. We present illustrative examples of the construction of some commonly used models along with the corresponding commands in Fig. 2. It should be noted that the model construction is performed in lattice coordinates. When the initial structure is non-orthogonal, this may lead to distortion of the constructed model.

### 1.2 Parameters for MD code interface

Since *MyTm* does not have built-in MD functionality, it relies on external code to perform the MD simulations. The external MD program is invoked by *MyTm* via a script, whose file name and interpreter are specified by ***md_script*** and ***md_script_interpreter*** in the *input.dat* file. For example, if the MD program is invoked via the bash interpreter and a script named *MyTm_MD.sh*, in the *input.dat* file, this is specified by setting:

***md_script ./MyTm_MD.sh***

***md_script_interpreter bash***

### 1.3 Temperature auto-adjustment parameters

In *MyTm*, temperature adjustment methods are realized by the ***run*** command specified in the *input.dat* file. The command for the continuous heating–until–melting procedure is shown below:

***run heatingup startTem endTem heatRatio***

Here, ***startTem*** and ***endTem*** are float numbers specifying an initial estimated temperature, with the aim of ensuring that the melting point lies between these two values. In practice, if the parameter ***init_tem_check*** is set to 1, *MyTm* automatically adjusts these values to ensure that the melting point falls within this range, although this may incur additional computational cost. The ***heatRatio*** float parameter specifies the heating rate. It should be noted that *MyTm* assumes a default MD timestep of 1 fs; therefore, if the user modifies the MD timestep, ***heatRatio*** must be scaled accordingly to maintain the correct heating rate.

For methods that use the bisection temperature adjustment, the command is

***run bisection startTem endTem breakType threshold***

Here, the parameters ***breakType*** is integer equal to 1 or 2, which serve as the stopping criteria for the bisection iterations. Specifically, ***breakType*** equal to 1 specifies the temperature interval threshold—when the temperature difference between the upper and lower bounds becomes smaller than ***threshold***, the bisection process terminates. The ***breakType*** equal to 2, that indicates the accuracy of the solid–liquid coexistence remains stable; In fact, *MyTm* computes the slope of the time-dependent solid fraction, and the iteration terminates when the slope falls below the value of parameter ***threshold.***

In addition, there are five parameters that control the details of temperature adjustment: ***max_steps***, ***max_sub_steps***, ***md_steps***, ***md_steps_adaption*** and ***bisection_relax_struct***. The ***max_steps*** parameter defines the maximum number of iterations for the bisection method. ***max_sub_steps*** controls the maximum number of repetitions for the MD simulation if equilibrium is not reached. The number of MD steps in each simulation is determined by ***md_steps***. The ***md_steps_adaption*** parameter is used to enable or disable equilibrium checks for the system. In fact, for the two-phase method with the NPT ensemble, it is not necessary for every MD simulation to reach full equilibrium, so setting ***md_steps_adaption*** to 0 can accelerate the calculation. For the implementations with the NPH ensemble, however, it is necessary to ensure that the system reaches an equilibrium solid–liquid coexistence state, thus ***md_steps_adaption*** should be set to 1. In principle, for ensembles without temperature control, where temperature equilibration relies solely on the dynamics, it is necessary to ensure that the solid-liquid model reaches equilibrium. Moreover, when ***md_steps_adaption*** is set to 1, ***md_steps*** will be automatically adjusted, which incurs additional computational cost. Therefore, when enabling ***md_steps_adaption***, it is strongly recommended to set a sufficiently large value of ***md_steps*** to reduce overall simulation time. The parameter ***bisection_relax_struct*** is used to control whether the solid–liquid model is re-optimized at each step of the bisection method, and it is recommended to open it (set adapt to 1) for relatively small models.

If the computational resource limitation restricts the initial model size to a relatively small system (typically on the order of a few hundred atoms), *MyTm* also provides the small-cell method[23]. In practice, this method does not constitute a direct simulation method. Therefore, it is implemented in *MyTm* only as a plugin. When the model is constructed as a solid–liquid coexistence system and the following command is used for temperature control, the small-cell method is activated:

***run smallcell startTem endTem***

It should be noted that, when the small-cell method is enabled, the parameters ***startTem*** and ***endTem*** will not be adjusted automatically. These parameters must be specified by the user, either based on prior knowledge or determined in advance using *MyTm* with other methods.

### 1.4 Solid-liquid atomic classification parameters

In *MyTm*, the solid–liquid identification is controlled by the parameter method specified in the input.dat file.

***method type radius_cutoff***

***type*** is an integer that specifies the structure identification method (1: SOP; 2: LAEO;

3: ML). ***radius_cutoff*** defines the cutoff radius for identifying neighboring atoms. When this value is less than 0, *MyTm* automatically uses the position of the first peak in the RDF as the cutoff radius. After *MyTm* distinguishes the atoms in the system, it calculates the fraction of atoms identified as solid. Two additional parameters, ***solid_cutoff*** and ***liquid_cutoff***, are then used to determine whether the whole system is classified as solid or liquid. If the fraction of solid atoms exceeds ***solid_cutoff***, *MyTm* considers the system as solid; if it is below ***liquid_cutoff***, the system is considered as liquid. This dual-threshold design introduces an adjustable tolerance region, allowing users to accommodate system-dependent thermal distortions while reducing the sensitivity of solid–liquid identification to the choice of a single critical threshold.

## 2. Initial model file

The initial model file is *POSCAR_init*, which adopts the structural format of VASP code. This file only requires the minimal repeating unit, which for crystals corresponds to the primitive cell. In fact, due to the limitations of solid-liquid classification methods, *MyTm* can only handle melting points of long-range ordered systems, and is not applicable to amorphous solids. For example, in the case of melting point calculations for Mg, the initial model file is shown in Fig. 4.

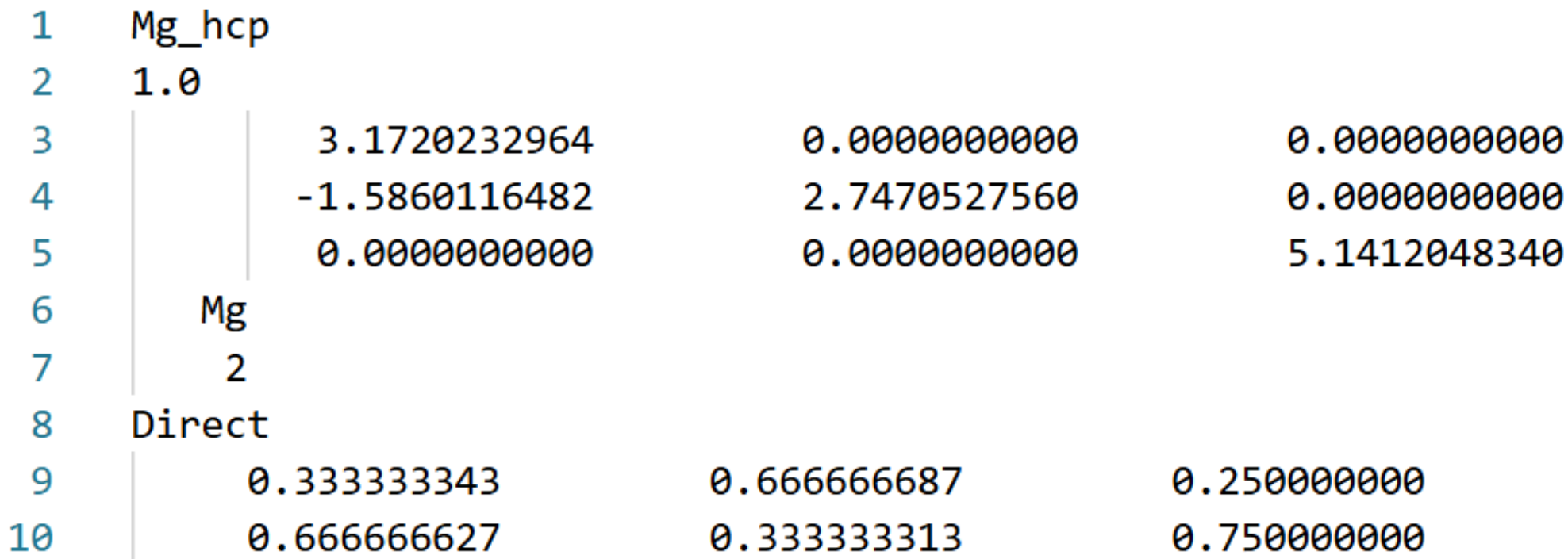

```
1   Mg_hcp
2   1.0
3          3.1720232964         0.0000000000         0.0000000000
4         -1.5860116482         2.7470527560         0.0000000000
5          0.0000000000         0.0000000000         5.1412048340
6      Mg
7       2
8   Direct
9        0.333333343         0.666666687         0.250000000
10       0.666666627         0.333333313         0.750000000
```

**Figure 4.** The model file format used in *MyTm*. It is in the same format as that of the VASP code.

## 3. MD code interface script

The main tasks of the MD interface script are as follows: (1). Perform a MD simulation. The initial temperature, final temperature, and total number of simulation steps are passed to the script as parameters (accessed within the script as ***$1, $2,*** and ***$3*** for bash script). The atomic structure model for the MD simulation is provided in the file *POSCAR_MyTm*. This file can be directly used when employing VASP code, while for other MD software, a format conversion may be required. (2). Post-processing after the MD simulation completes. The thermodynamic quantities obtained during the simulation—temperature, pressure, volume, and atomic trajectories—should be written to the files *Temperature_MyTm*, *Pressure_MyTm*, *Volume_MyTm, and XDATCAR_MyTm*, respectively. The trajectory file follows the same format as that of VASP.

It should be noted that *MyTm* is unable to detect error messages returned by external MD programs. As a result, it passes additional parameters corresponding to the current melting point calculation step (accessible as ***$4*** and ***$5***), which correspond to ***max_step*** and ***max_sub_step*** in the *input.dat* file. Users may optionally back-up the MD simulation results for later inspection of the melting point calculation process and verify its consistency. A typical example of a *MyTm_MD.sh* script for VASP is shown in Fig. 5.

```
1   #!/bin/bash
2
3   # get parameters for MD
4   tem_start=$1
5   tem_end=$2
6   nsw=$3
7   steps=$4
8   sub_steps=$5
9
10  # run MD by VASP code
11  cp INCAR_init INCAR
12  cp POSCAR_MyTm POSCAR
13  sed -i "1i TEBEG=${tem_start}" ./INCAR
14  sed -i "1i TEEND=${tem_end}" ./INCAR
15  sed -i "1i NSW=${nsw}" ./INCAR
16  mpirun -n 32 vasp6.4_gam > vasp.log
17
18  #get MD trajectory, temperature, pressure and volume to file
19  cp XDATCAR XDATCAR_MyTm
20  grep 'T=' OSZICAR |awk '{print NR,$3}' > Temperature_MyTm
21  grep 'volume of cell :' OUTCAR |awk '{print NR,$5}' > Volume_MyTm
22  grep 'total pressure  =' OUTCAR |awk '{print NR,$4}' > Pressure_MyTm
23
24  #MD result backup
25  mv XDATCAR XDATCAR_${steps}_${sub_steps}
26  mv OUTCAR OUTCAR_${steps}_${sub_steps}
27  mv OSZICAR OSZICAR_${steps}_${sub_steps}
```

**Figure 5.** Example script file for *MyTm* calling external MD software (here VASP). Lines 3–8: Retrieve the MD simulation temperature, number of steps, and *MyTm* steps. Lines 10–16: Run the molecular dynamics simulation. Lines 18–22: Extract the MD simulation trajectories of temperature, pressure, and volume into corresponding files. Lines 24–27: Backup the MD simulation results.

## 4. Melting point calculation

After preparing all input files, run the program again with the parameter "***-r***". *MyTm* will automatically complete model construction and melting point calculation. The results of melting temperature calculation are written to *MyTm_result.dat* file, which mainly consists of five sections: input parameters, temperature adjustments at each step, thermodynamic information, atomic solidification ratios and the final calculated melting point. The information in *MyTm_result.dat* file is shown in Fig. 6. Detailed information regarding all parameters and files is provided in Tables 1 and 2.

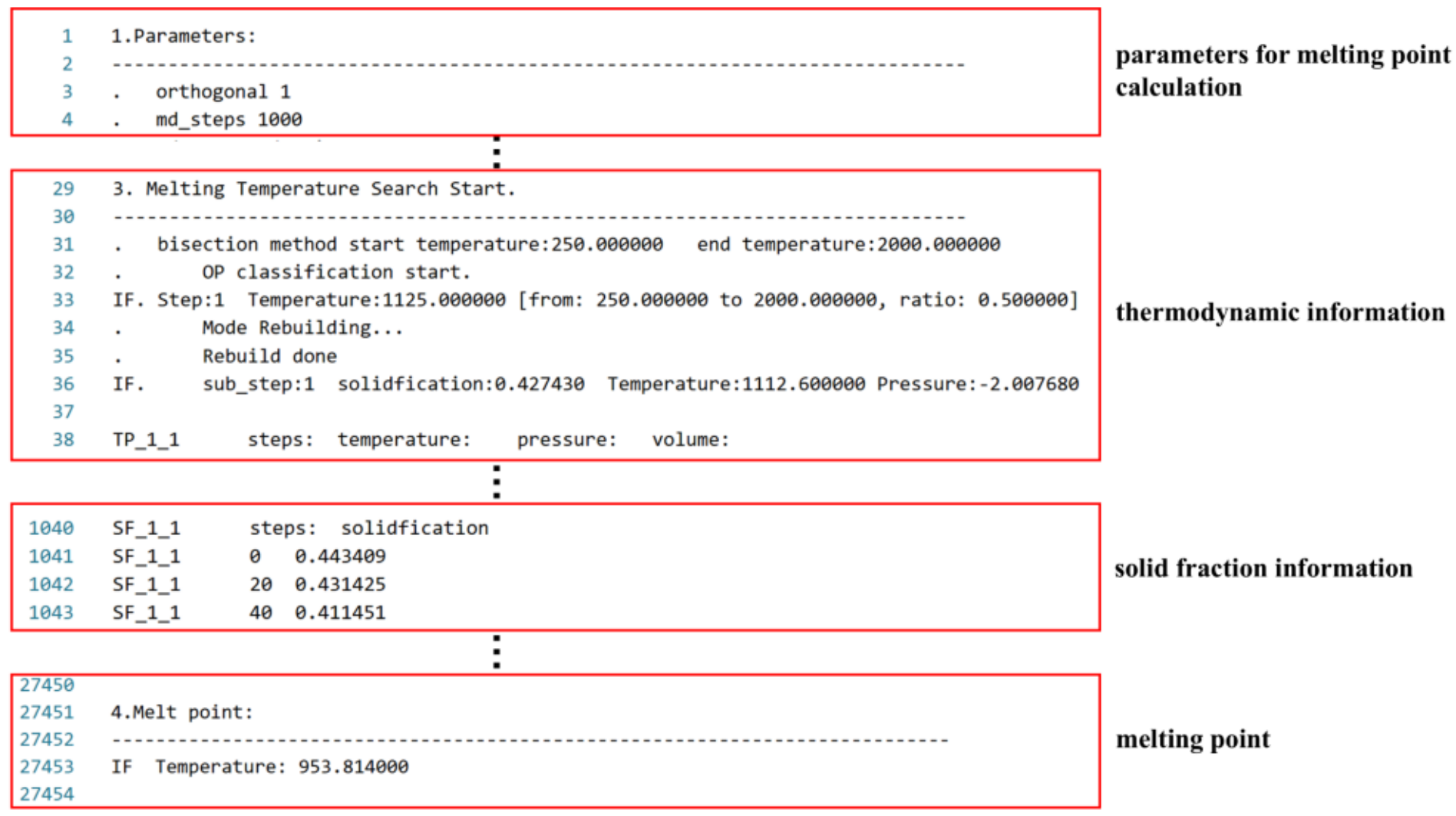

**Figure 6.** File format of *MyTm_result.dat* files. The output file mainly includes input parameters, temperature adjustment information, thermodynamic data from MD simulations, solidification ratios, and the melting point.

**Table 1.** Input and output files.

| File name | **description** |
|---|---|
| *input.dat* | The control file for *MyTm*; can be automatically generated using the command ***MyTm -m.*** |
| *POSCAR_init* | Initial model (in VASP structure file format.) |
| *Temperature_MyTm* | Temporary file, storing the temperature of MD. |
| *Pressure_MyTm* | Temporary file, storing the pressure of MD. |
| *Volume_MyTm* | Temporary file, storing the cell volume of MD. |
| *XDATCAR_MyTm* | Temporary file, storing the trajectory of MD. |
| *POSCAR_MyTm* | Temporary file, Model for Molecular Dynamics. |
| *MyTm_ml_train.data* | When ML is enabled to recognize atoms, the network structure, training process, and result validation data are saved to this file. |
| *MyTm_result.dat* | All output information file of *MyTm.* |

**Table 2.** Control parameters of *MyTm*.

| parameter name (module) | value type (default) | description |
|---|---|---|
| ***Method*** (atomic classification) | Integer (0) | Classification method for solid and liquid atoms.<br>1 represents SOP classification<br>2 represents LAEO classification<br>3 represents ML classification |
| ***liquid_cutoff*** (atomic classification) | Float (0.8) | Cutoff value for liquid determination. |
| ***solid_cutoff*** (atomic classification) | Float (0.2) | Cutoff value for solid determination. |
| ***md_steps*** (MD interface) | Integer (1000) | The number of steps in MD simulation. |
| ***md_steps_adaption*** (MD interface) | Integer (0) | Whether to automatically adjust the number of MD simulation steps.<br>0 represents No.<br>1 represents YES |
| ***md_script*** (MD interface) | String (None) | The script path and name for *MyTm* to call MD. |
| ***md_script_ interpreter*** (MD interface) | String (None) | The interpreter name for script, such as python, bash, etc. |
| ***supercell*** (model) | Integer (1 1 1) | The multiplier to expand the MD simulation box along the first lattice vector. |
| ***build*** (model) | Command (None) | Constructing defects or coexisting solid-liquid phases.<br><br>***build type shape1 shape2 defect_scale***<br><br>***type***=***defect*** or ***liquid.*** Create defect areas or liquid phase<br><br>***shape1***/***shape2*** =**0, 1, 2, 3**. Create geometric shapes for liquid phase regions or defect regions. ***shape1*** represents the interior of geometric shapes, while ***shape2*** represents the exterior. (0: None, 1: Parallelepiped shape, 2: Ellipsoid shape, 3: Cylinder shape)<br><br>***defect_scale*** represents the proportion of atoms removed from defects. |
| ***size*** (model) | Command (None) | The size of the geometric region in the build command<br><br>***size a1 b1 c1 a2 b2 c2***<br><br>***a1, b1, c1*** represent the size of the parallelepiped of ***shape1***.<br><br>***a2, b2, c2*** represent the size of the parallelepiped of ***shape2***. |
| ***pos*** (model) | Command (None) | Center position of defect or liquid phase<br><br>***Pos x1 y1 z1 x2 y2 z2*** |
| ***rot*** (model) | Command (None) | Rotation angle of defect or liquid phase along x, y and z axis.<br><br>***rot x1 y1 z1 x2 y2 z2*** |
| ***max_steps*** (temperature adjustment) | Integer (20) | Maximum number of iterations for temperature adjustment. |
| ***max_sub_steps*** (temperature adjustment) | Integer (40) | The maximum number of repeated calls to MD simulation in a single iteration. |
| ***run*** (temperature adjustment) | Command (None) | Temperature adjustment method.<br><br>***run heatingup startTem endTem heatRatio***<br><br>***startTem*** and ***endTem*** defines the initial temperature control interval of ***heatingup, bisection*** and ***smallcell***. ***heatRatio*** defines the heating rate of ***heatingup.***<br><br>***run bisection startTem endTem breakType threshold***<br><br>***breakType*** equal to 1 defines the temperature convergence and ***breakType*** equal to 2 defines solid-liquid ratio stability accuracy of ***bisection.*** Parameter ***threshold*** defines convergence tolerance.<br><br>***run smallcell startTem endTem*** |

# IV. Validation

In this section, we selected a variety of systems to test *MyTm*. First, we evaluated *MyTm* using three different atomic classification methods (SOP, LAEO, and ML) on solid–liquid coexistence structures, and the results are shown in Fig. 7. As shown in Fig. 7(a), the conventional methods based on atomic density and diffusion coefficients perform relatively poorly in identifying solid–liquid mixtures, with misclassifications occurring in both solid and liquid regions. In contrast, within *MyTm*, the SOP and LAEO methods show significantly improved solid–liquid identification performance, with only minor misclassifications observed in a few systems for both solid and liquid phases (e.g., Mg, Cu, and NaCl). The ML-based method exhibits almost no misclassification, indicating a high level of robustness for atomic phase identification across different systems.

(a) conventional classification comparison
Solid
Liquid
density (Å$^{-3}$)
Density classification
Statistical curve of atomic diffusion coefficient
The number of atoms
Solid
Liquid
cutoff value of D
D (diffusion coefficient) (Å$^2$/fs)
Diffusion Coefficient classification
(b) *MyTm* classification comparison
Mg
Fe
Cu
NaCl
NiAl
SOP
LAEO
ML

Figure 7. (a) Identification results in a solid–liquid coexistence model using the conventional atomic density method and the diffusion coefficient method. (b) Results identified by the SOP, LAEO, and ML methods in *MyTm*. For clarity, atoms identified as belonging to the solid phase are uniformly colored yellow, while those identified as belonging to the liquid phase are colored green.

We illustrate the automatic evaluation of the melting points for several representative materials (Mg, MgO, Na, NaCl, Fe, NiAl) using *MyTm*. There is no human intervention involved for all of these for calculations. The results obtained from various melting point determination methods as implemented in *MyTm* are compared with previously reported theoretical evaluation value[58–63] and experimental data[64–67] in Table 3. We can see that the melting points calculated using both the solid–liquid coexistence method and the modified void method implemented in *MyTm* are nearly identical to those reported in the literature, which demonstrates the practical applicability and accuracy of *MyTm*. In contrast, the melting points obtained using the direct-heating method and the void method are generally overestimated due to their inherent limitations in line with the current understanding about these methods. In fact, they are usually recommended only as an upper bound estimation for the melting temperature or for rapid estimates of melting in the literature.

**Table 3.** Melting point calculated by *MyTm*. All temperatures are given in Kelvin (K)

| | Mg | Fe | Cu | MgO | NaCl | NiAl |
|---|---|---|---|---|---|---|
| Direct-heating method | 1281 | 2281 | 1728 | 3906 | 1468 | 2656 |
| Void method | 1156 | 2031 | 1656 | 3593 | 1468 | 2531 |
| Modified Void method | 1011 | 1825 | 1219 | 3190 | 1041 | 1856 |
| Solid-liquid coexistence method | 914 | 1777 | 1356 | 3398 | 1072 | 1828 |
| Sandwich method | 914 | 1621 | 1328 | 3146 | 1015 | 1857 |
| Z method | 1207 | 2149 | 1632 | 3651 | 1397 | 1815 |
| Modified Z method | 944 | 2246 | 1308 | 3147 | 1025 | 1825 |
| Ref. | 918 | 1750 | 1356 | 3295 | 1061 | 1850 |
| Exp. | 923 | 1811 | 1358 | 3125 | 1074 | 1911 |

## Conclusions

In this work, we constructed a melting-point calculation pipeline workflow that consists of four components—model construction, ensemble selection, temperature control, and solid–liquid atom classification—and successfully implemented in a fully automated melting-point calculation program. It resolves the notorious challenges in direct MD simulation of melting that requires heavy human interventions. The flexible MD code interface and automatic calculation features make *MyTm* an excellent choice for high-throughput melting-point computing. Moreover, the built-in lightweight MLP classification effectively resolves the issues of misclassification and low accuracy that commonly encountered in traditional solid–liquid atomic identification, thereby making *MyTm* very robust and reliable for melting-point determination.

## Declaration of competing interest

The authors declare that they have no known competing financial interests or personal relationships that could have appeared to influence the work reported in this paper.

## Data availability

Data will be made available on request.

## Acknowledgements

This work was supported by the National Key R&D Program of China under Grant No. 2021YFB3802300, the National Natural Science Foundation of China under Grant No. 12404287, 12372370. Part of the computation was performed using the supercomputer at the Center for Computational Materials Science (CCMS) of the Institute for Materials Research (IMR) at Tohoku University, Japan.

## CRediT authorship contribution statement

Yu S. Huang: Investigation, Methodology, Writing - original draft, Writing - review & editing. Hong X. Song: Methodology, Writing - review & editing, Writing - review & editing. Y. Sun: Methodology, Writing - review & editing, Writing - review & editing. Jin L. Li: Methodology. Yin L. Xu: Methodology. F. C. Wu: Methodology. Y. C. Gan: Methodology. Yu. F. Wang: Methodology. Hao Wang: Methodology, Writing - review & editing. Hua Y. Geng: Conceptualization, Project design, Writing, Reviewing, and Editing, Supervision, Project administration, Software.